\documentclass{svproc}
\usepackage{graphicx}
\usepackage{booktabs}
\usepackage{xcolor}
\usepackage{amssymb}
\usepackage{amsmath} 

\definecolor{lighter_red}{RGB}{255, 204, 204}
\definecolor{darker_red}{RGB}{204, 0, 0}

\usepackage{url}

\begin{document}
\mainmatter              
\title{Turbulent/non-turbulent interface in high Reynolds number pressure gradient boundary layers}
\titlerunning{APG}  
%
\author{Ivan Marusic\inst{1} \and Wagih Abu Rowin\inst{1} \and
Mitchell Lozier\inst{1} \and Luka Lindić\inst{1} \and Ahmad Zarei\inst{1} \and Rahul Deshpande\inst{1}}
\authorrunning{Marusic et al.} 
%
%
\institute{$^{1}$Department of Mechanical Engineering, University of Melbourne, Parkville, VIC 3010, Australia \\
\email{imarusic@unimelb.edu.au},\\
home page: \texttt{https://people.eng.unimelb.edu.au/imarusic/index.html/}}

\maketitle             

\begin{abstract}
We report two-dimensional particle image velocimetry experiments in high Reynolds number turbulent boundary layers imposed with a moderately strong streamwise pressure gradient.
The unique aspect of these data are the highly resolved measurements across the outer region of a physically thick boundary layer, enabling accurate detection of turbulent/non-turbulent interfaces (TNTI).
The present dataset is used to detect the TNTI of an adverse pressure gradient turbulent boundary layer and compare its characteristics with that for a zero-pressure gradient boundary layer, at a nominally similar friction Reynolds number.
It is found that the TNTI exists across a broader range of wall-normal distance in presence of an adverse pressure gradient, as compared to the  zero-pressure gradient case.
Implications on conditionally averaged turbulence statistics are discussed based on detection of the TNTI.

\keywords{Experimental Fluid Mechanics, Wall-bounded turbulence, Adverse Pressure Gradient}
\end{abstract}
\section{Introduction}
A characteristic feature of any turbulent flow is a broad spectrum of turbulent scales of motion, or eddies, whose dynamics are governed by the viscous and inertial forces acting on the flow.
In turbulent flows, the relationship between these viscous and inertial forces has traditionally been described by the Reynolds number, a parameter which is non-dimensionalized using a characteristic length scale, velocity scale and viscosity, which are specific to each flow. 
These forces, and consequently the dynamics of the turbulent eddies, are confined within and regulated by the boundaries of the turbulent fluid flow.
In this study, we will limit our focus to wall-bounded turbulent flows, specifically the turbulent boundary layer (TBL), which is technologically relevant to many engineering applications such as over aircraft wings and ship hulls.

In the case of the TBL, one boundary is a solid wall (with discussion here limited to smooth walls only), while the second boundary is a complex and freely developing interface between the turbulent eddies within the TBL (turbulent region) and the external freestream flow (non-turbulent region).
This interface will be referred to as the turbulent/non-turbulent interface (TNTI) henceforth \cite{daSilva2014}.
In the case of the TBL, we will refer to the friction Reynolds number $Re_{\tau}=U_{\tau}\delta/\nu$, where $U_{\tau}$ and $\nu$ denote the mean friction velocity and kinematic viscosity respectively. 
We limit our discussion here to nominally high-$Re_{\tau}$ ($\gtrsim$ 7000) TBLs, which have only been investigated experimentally to date.
There are several methodologies reported in the literature on estimating the boundary layer thickness, $\delta$; all of them essentially indicate the local wall-normal distance demarcating the freestream flow from the turbulent region, in a time-averaged sense.
The freestream flow is considered to be irrotational and non-turbulent, although in practice there is almost always some measurable turbulence in the freestream flow.
This freestream turbulence intensity varies depending on the experimental facility, but is typically orders of magnitude lower than the intensity of turbulent eddies within the TBL, such that the two regions are distinguishable.
In addition to spatially identifying the turbulent and non-turbulent regions, the dynamics of this bounding TNTI also regulates the transport of important quantities such as mass, momentum, and energy between the turbulent and non-turbulent regions of the flow \cite{buxton_entrainment_2023}.
As such the mixing and entrainment processes associated with the TNTI are not only interesting, but also pertinent to many engineering applications, and are an active area of research for a variety of turbulent flows.

Many methods and measurement techniques have been used to locate or describe the TNTI of the TBL.
Early methods utilized time-resolved measurements of the streamwise velocity from hot-wires to generate a statistical description of the TNTI.
In the wake region of the TBL, large-scale pockets or bulges containing turbulent eddies which passed the hot-wire were highly contrasted with the relatively quiescent freestream flow creating a highly intermittent velocity signal.
The turbulent parts of the signal were distinct due to the presence of high frequency and high amplitude turbulent fluctuations.
This aspect of the velocity signal can be enhanced by differentiating \cite{hedley_decisions_1974} or high-pass filtering \cite{fransson_transition_2005} the velocity signal, creating a so-called detector function.
By applying a threshold to this detector function the turbulent portions of the signal can be identified and used to quantify the wake-region intermittency or compute conditionally averaged statistics for example \cite{krug_2017}.
More recently, spectral methods such as wavelet analysis \cite{de_wavelet_2023}, have also been employed to achieve similar outcomes.
The spectral signature of the high frequency and high amplitude fluctuations, associated with the turbulent regions of the flow, are easy to identify and a detector type function can be created by averaging the spectral energy of relevant eddies.
Each of these methods has some inherent subjectivity stemming from the use of thresholds and/or filtering/smoothing techniques which are applied to the velocity signals and/or detector functions \cite{de_wavelet_2023,hedley_decisions_1974}.
With the rise of spatially resolved two-dimensional velocity measurement techniques (e.g. particle imaging velocimetry, PIV), new methods were used to find a two-dimensional instantaneous representation of the TNTI.
One method is introducing passive tracer particles into the turbulent and/or non-turbulent regions of the flow in order to make them distinguishable during imaging.
However, quantities such as the vorticity \cite{semin_superlayer_2011} or local kinetic energy \cite{chauhan_entrainment_2014} have also been used successfully to delineate the turbulent and non-turbulent regions of the TBL without additional tracers.
With these methods, once the TNTI is identified, various properties associated with the instantaneous TNTI can be investigated such as mass/energy flux, multi-scale dynamics, or conditional statistics \cite{chauhan_entrainment_2014}.
Similar methods have also been applied with three-dimensional measurement techniques.

A majority of the previous research effort regarding TBL TNTI characteristics has been focused on canonical (i.e. smooth-wall, zero-pressure gradient (ZPG)) TBLs.
However, TBLs encountered in engineering applications are typically exposed to non-canonical effects such as pressure-gradients.
In particular, adverse-pressure gradients (APG) have been shown to affect the dynamics of large-scale eddies, one consequence of which is an increase in the growth rate of the boundary layer thickness \cite{harun2013pressure}.
It would follow that APG conditions can have an affect on the characteristics and dynamics of the TNTI as well \cite{yang2020turbulent}.
In order to study TNTI dynamics of an APG TBL at high $Re_{\tau}$, we will follow the technique described by Chauhan et al. \cite{chauhan_entrainment_2014} utilizing the local kinetic energy (LKE), measured using well-resolved planar PIV in the wall-normal direction, to locate the instantaneous TNTI bounding the TBL.
To the best of the author's knowledge, this work represents the first analysis of TNTI employing the LKE methodology for a high-$Re_{\tau}$ turbulent boundary layer with an adverse-pressure gradient.

As per Chauhan et al. \cite{chauhan_entrainment_2014}, the local kinetic energy, $\tilde{k}$, over a $3 \times 3$ spatial grid can be computed via equation \ref{LKE}.

\begin{equation}
    \tilde{k}=100 \times \frac{1}{9 U_{\infty}^2} \sum_{m, n=-1}^1\left[\left(\tilde{U}_{m, n}-U_{\infty}\right)^2+\left(\tilde{W}_{m, n}\right)^2\right]
    \label{LKE}
\end{equation}


Here the instantaneous streamwise and wall-normal velocity components are denoted by $\tilde{U}$ and $\tilde{W}$ respectively, with $U_{\infty}$ representing the mean freestream velocity.
The corresponding streamwise and wall-normal axes will be represented by $x$ and $z$ respectively.
In the region above the TNTI, it is expected that $\tilde{k}$ will be equal to the freestream turbulence intensity ($\overline{u^2}/U^{2}_{\infty}$ which is facility dependent), while below the interface $\tilde{k}$ will increase significantly approaching the wall.
As such an LKE threshold, $k_{th}$, is chosen which corresponds to the known freestream turbulence intensity of the facility.
A binary representation of the instantaneous flow field can then be obtained where non-turbulent regions ($\tilde{k}<k_{th}$) are assigned a value of $1$ and turbulent regions ($\tilde{k} \gtrsim k_{th}$) are assigned a value of $0$.
The local interface position is then extracted using a contour algorithm to generate a contour line at a level of $0.5$ which demarcates the turbulent and non-turbulent regions.
With the instantaneous TNTI identified, a range of statistical calculations can then be undertaken to study the TNTI dynamics.
This paper reports preliminary analysis of the conditional statistics based on TNTI detection, for both ZPG and APG TBLs.

\section{Experimental setup}

\begin{figure}[t!]
\centering
\includegraphics[width=\linewidth]{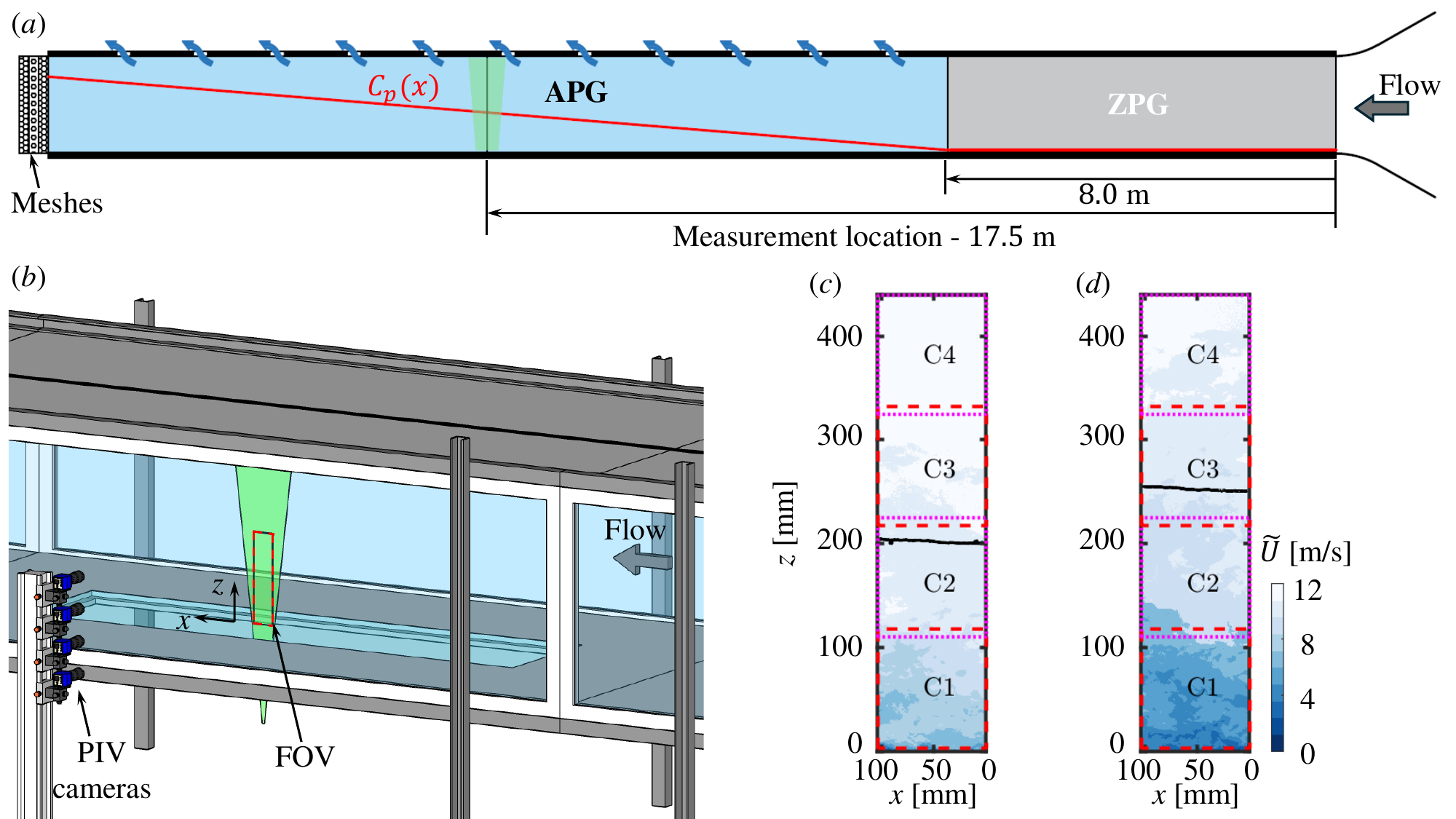}
\caption{(\textit{a}) Schematic of the Melbourne wind tunnel modified to adjust the static-pressure coefficient ($C_p$) as a function of the streamwise distance, $x$. The arrows on the test section ceiling represent air bleeding on introduction of the blockage at the tunnel outlet (via meshes/screens). (\textit{b}) Schematic illustrating the planar PIV setup, featuring the tower configuration of four cameras and the laser sheet passing through the glass tunnel floor.  
Instantaneous streamwise velocity, $\tilde{U}$-field for the (\textit{c}) ZPG and (\textit{d}) APG cases in the $xz-$plane. 
In (\textit{c}) and (\textit{d}), the flow direction is from right to left, and the boundary layer thickness (${\delta}_{99}$) is estimated based on the methodology used by Bobke et al. \cite{bobke2017history} and is indicated by black lines. The red and magenta rectangles in (\textit{c}) and (\textit{d}) represent the captured field-of-view by each camera, with each view labelled with the respective camera number, e.g., C1, C2.}
\label{APG3setup}
\end{figure}

\subsection{Facility and APG setup}
Both the ZPG and APG flow scenarios were established in the large Melbourne wind tunnel facility (HRNBLWT; \cite{marusic2015evolution,deshpande2023reynolds}), which permits precise control of the pressure gradient via air bleeds. 
The air bleeds are essentially spanwise slots existing on the tunnel ceiling, that are strategically spaced every 1.2\:m along the 27\:m long working section of the HRNBLWT. 
Figure \ref{APG3setup}(a) shows a schematic of the experimental setup, which includes a custom-designed frame installed at the test section outlet (inspired from Clauser \cite{clauser1954turbulent}), that holds multiple screens/wire meshes to enforce the pressure gradient in the test section. 
These screens are made out of perforated steel, with a porosity of 51\%. 
The magnitude of the adverse pressure gradient can be easily changed by varying the number of screens \cite{deshpande2023reynolds}.
A ZPG condition is imposed across the test section by removing all the low porosity ($\sim$51\%) screens and installing a single high porosity screen at the outlet, similar to the original configuration of the HRNBLWT reported in Marusic et al. \cite{marusic2015evolution}. 

The unique aspect of the present wind tunnel setup is its ability to impose an APG at a sufficiently downstream location, after the TBL has grown under a nominally ZPG condition upstream (\emph{i.e.}, a canonical condition). 
This is schematically indicated in figure \ref{APG3setup}(a), which shows a well-established canonical inflow ZPG up to $x \approx 8 m$, maintained via choking the air bleeding from the tunnel ceiling. 
The APG is only enforced in the downstream section of the tunnel by permitting air bleeding from the ceiling, which leads to a quasi-linear growth in the tunnel static pressure (indicated by non-dimensional pressure coefficient, $C_p$($x$) in figure \ref{APG3setup}a).
Here, the definiton of $C_p$($x$) is based on the freestream velocity variation along $x$ (\emph{i.e.}, ${U_{\infty}}(\rm x)$), given by:
\begin{equation}
{C_p}(\rm x) = 1 - {{U^2_{\infty}}(\rm x)}/{{U^2_{\infty}}(\rm x = 0)}   
\end{equation}
The $C_p$ variation obtained in this facility is unique with respect to the vast majority of past studies on APG TBL flows, which have predominantly relied on working-section inserts to produce a varying cross-sectional area to establish the adverse pressure gradient \cite{knopp2021experimental,song2002effects,bross2019interaction}. 
These working section inserts usually involve a constriction, leading to a favourable-pressure gradient, followed by a limited straight section for a zero-pressure gradient, before finally leading to the APG region. 
While each of these previous studies is in itself interesting, every change/perturbation in the upstream PG condition is known to result in unique history effects that affect the flow downstream \cite{bobke2017history,deshpande2023reynolds}. 
In the select few past experiments without inserts \cite{perry1995wall,monty2011parametric,harun2013pressure}, a canonical ZPG boundary layer led into an APG region, which was found to minimize the upstream history effects. 
However, all these experiments were conducted at $Re_\tau$ $\lesssim$ 3 000, and thus restricted to low-moderate Reynolds numbers.
For this purpose, the present study will investigate APG TBL flow at high $Re_{\tau}$ ($\sim$7000), which will be compared with ZPG case at nominally matched $Re_\tau$.
Interested readers can find more details on this experimental setup in Deshpande et al. \cite{deshpande2023reynolds}.

This study quantifies the intensity of the pressure gradient by defining the Clauser pressure gradient parameter, $\beta$ \cite{clauser1954turbulent} defined as:
\begin{equation}
{\beta} = (\delta^* / {\rho {{U}}_\tau^2}) \left( {\rm d} {P}/ {\rm d} x \right), 
\end{equation}
where $\delta^* = \int_0^{\delta_{99}}(1 - {\overline{U}}(z)/{{U_e}})\:{\rm d}z$ is the displacement thickness,   
$U_e$ $=$ $\overline{U}$($z$ = $\delta_{99}$), \textit{i.e.} the mean streamwise velocity at the edge of the TBL, $\rho$ is the fluid density and ${\rm d} {P}/{\rm d} x$ is the mean streamwise pressure-gradient at the $x$-location where $\beta$ is estimated.
All the experimental cases considered in this study, and their associated measurement conditions, have been documented in table \ref{details}.
Here, $\delta_{99}$ is quantified using the same definition as used by Bobke et al. \cite{bobke2017history} and Deshpande et al. \cite{deshpande2023reynolds}, based on the diagnostic plot method.
Both $\delta_{99}$ and ${\delta}^*$ are quantified for the ZPG and APG cases via the PIV data, which will be described next ($\S$\ref{piv}).
Friction velocity, $U_{\tau}$ has been estimated directly via oil-film interferometry (OFI), which will be described in $\S$\ref{ofi}.



\subsection{Planar PIV setup} 
\label{piv}

To acquire detailed measurements across the entire boundary layer, we employed four high-resolution Imager CX-25 12-bit resolution cameras staggered vertically to capture a narrow vertical column spanning the entire boundary layer height, as shown in figure~\ref{APG3setup}(\textit{b}). 
Each camera features a complementary metal oxide semiconductor (CMOS) sensor with dimensions of ${5312\times 4608}$ pixels. The CMOS sensor possesses pixels measuring $2.7 \times 2.7$~\textmu m\textsuperscript{2}. Equipped with Tamron SP AF 180 mm macro photography lenses  set at an aperture size of $f/11$, each camera achieved a digital resolution of 22~\textmu m/pixel. Illumination was provided by an InnoLas SpitLight Compact PIV 400 dual pulse Nd:YAG laser, with typical spherical and cylindrical lenses configuration. To illuminate the boundary layer with minimum reflection in the vicinity of the wall, a laser sheet with an estimated thickness of 2~mm was directed from the bottom glass as illustrated in figure~\ref{APG3setup}(\textit{b}). 
Both the PIV measurements, in a high $Re_{\tau}$ ZPG and APG TBL, were conducted at the same streamwise location, $x$ $\sim$ 17.5\:m (figure \ref{APG3setup}a). 
The flow was seeded with polyamide particles ranging from 1–2 µm in diameter. The final stitched field-of-view (FOV) measured $104 \times 441$~mm\textsuperscript{2} (in $x \times z$).

\begin{table}[t!]
  \centering
  \caption{Details of the PIV data sets acquired for ZPG and APG scenarios}
  \begin{tabular}{cccccccccc}
    \toprule
            Case & Symbol & $x$ & $U_{\infty}(x=0)$ & $U_{\infty}(x)$ & $Re_\tau$ & $\frac{\nu}{u_\tau}$ & $\delta_{99}$ & $\beta{(x)}$ & {\rotatebox[origin=c]{90}{\#PIV Images}} \\
    & & (m) & (m/s) & (m/s) & & (\textmu m) & (mm) & & \\
    \midrule
    ZPG &\textcolor{lighter_red}{$\Diamond$} &17.5 &12.6 &12.6 &7500 &36 &270 &0 & 2000\\
    APG &\textcolor{darker_red}{$\Diamond$}   &17.5 &12.6 &11.2 & 7100&46 &328 &1.4 & 2000\\
    \bottomrule
  \end{tabular}
  \label{details}
\end{table}

The synchronization of laser pulses and the four high-speed cameras was achieved using a programmable timing unit (PTU X, LaVision GmbH) controlled by DaVis~{10.1}. An ensemble of 2000 pairs of double-frame images with a laser pulse separation of 50~\textmu s was recorded. This pulse separation resulted in a maximum particle displacement of 10~pixels in the freestream region of $U_\infty = 12.6$~m\,s\textsuperscript{-1}. For camera calibration, a 2-D target comprising of 1~mm diameter dots, evenly spaced 5~mm apart, was utilised. 
The signal-to-noise ratio of the images was enhanced by subtracting the minimum intensity from the ensemble and subsequently normalizing them using the average intensity. Multi-pass cross-correlation was then applied with a final interrogation window size of $24\times 24$~pixels ($0.53 \times 0.53$~mm\textsuperscript{2}) with 50\% overlap. The final window sizes, along with the laser sheet thickness, correspond to viscous-scaled spatial resolutions in the $x^+ \times y^+ \times z^+$ directions of {$18 \times 55 \times 18$} and {$11 \times 43 \times 11$} for the ZPG and APG cases, respectively, where the spanwise spatial average is associated with the 2~mm thick laser sheet. 
Here the superscript `+' denotes normalisation with the friction velocity $U_\tau$ and kinematic viscosity $\nu$.  

As can be noted above, the unique aspect of current PIV data set is its excellent wall-normal resolution $\sim$10-20 viscous units across the entire TBL, making it ideal for detecting and analyzing TNTI interfaces.
Representative snapshots of the instantaneous streamwise velocity $U$ in the $x \times z$ plane are depicted in figures~\ref{APG3setup}(\textit{c}) and (\textit{d}) for the ZPG and APG TBLs, respectively. The annotated FOVs for each camera demonstrate the advantageous positioning of the current camera setup. 
In the ZPG case, the BL thickness, estimated based on ${\delta}_{99}$ lies in the FOV of camera 2. 
On the other hand, $\delta_{99}$ for APG TBL is situated close to the center of camera 3. This arrangement of cameras ensures high-resolution velocity measurement coverage of the entire BL for both the ZPG and APG experiments.

\subsection{Oil-film interferometry}
\label{ofi}

Oil film interferometry (OFI) is a direct and independent method for obtaining exact measurement of $U_{\tau}$ in turbulent boundary layers, especially at high Reynolds number where near-wall measurements are often inaccessible \cite{henry2007}.
This friction velocity measurement technique has emerged as an important requirement for testing of scaling arguments in high Reynolds number TBLs, such as the validity of the inner-scaling, log-region, etc.

\begin{figure}[t!]
\centering
\includegraphics[width=1.0\linewidth]{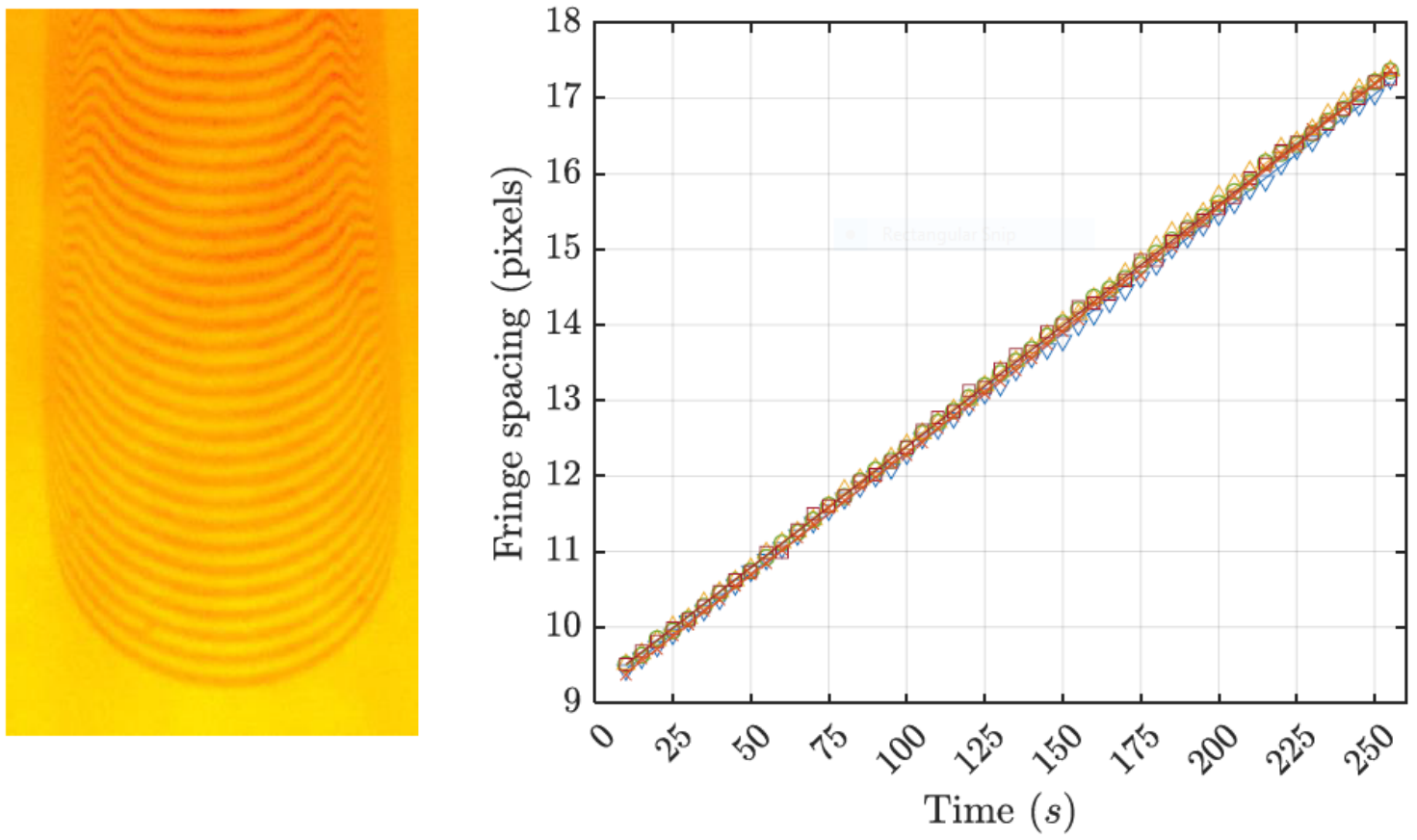}
\caption{Sample interferogram pattern and oil film fringe spacing for APG case}
\label{fig2}
\end{figure}

During an OFI experiment, a silicone oil (50 cSt) droplet is spread onto a smooth clean surface of the wind tunnel. 
As the droplet spreads into a thin film during `wind on' condition, it is illuminated by a monochromatic light source, creating visible fringes captured by a DSLR camera (depicted in figure \ref{fig2}a). 
The rate of change of these fringes correlates with the thinning of the oil film, which relates to the local wall shear stress. 
Using a Nikon D800 DSLR camera in time-lapse mode, over 100 images are taken in a 5-second interval for each OFI database. 
These images are processed using a fast-Fourier-transform algorithm to extract fringe spacing from interferograms. As explored by Fernholz et al. \cite{fernholz1996}, the wall shear stress can be expressed as the distance between fringes (${\Delta}x$) over a period of time (${\Delta}t$) as:
\begin{equation}
{\tau_w} = {\mu_{oil}}{\frac{{\Delta}x}{{\Delta}t}}{\frac{\sqrt{  {n^2_{oil} -- {n^2_{air}}{sin{\theta}} }}}{\lambda}}
\end{equation}
where $n_{air}$  and $n_{oil}$  denote the refractive indices of air and oil, respectively. 
The $\theta$  and $\lambda$ respectively  represent the illumination incident angle and wavelength of the light source (equal to 589.9\:nm for the sodium lamp utilized). 
An example of oil droplet fringe patterns and evolution of the fringe spacing across an OFI test are shown on the left and right side of figure \ref{fig2}. 
The linear fringe spacing over time shows the quality of fringe patterns, and consequently the OFI results. 
Considering uncertainties associated with oil viscosity calibration, fringe extraction, and potential dust contamination of the oil film, the uncertainty obtained by OFI in this study is estimated to be within $\pm$1.5\%.
The OFI experiment is conducted at $x$ $\sim$ 17.5\:m of the test section, for both ZPG and APG TBL experiments, to obtain their respective $U_{\tau}$ estimates at the exact same location as the PIV.

\section{Results}
This section presents and discusses results from our high-resolution PIV measurements for both the ZPG and APG cases, with a particular focus on estimating and comparing the TNTI characteristics for the two cases. 
Initially in \textsection{\ref{sec:vel_profiles}}, the mean statistics obtained from the PIV measurements will be validated by comparing them with the hot-wire data of Deshpande\textit{ et al}.\cite{deshpande2023reynolds}, which were obtained under identical flow conditions as the present data (i.e. $Re_{\tau}$, $\beta$ and $x$-location). 
Next, in \textsection{\ref{sec:TNTI_detect}}, we deploy the methodology proposed by Chauhan et al. \cite{chauhan_entrainment_2014} to detect the TNTI for both ZPG and APG cases. 
Finally in \textsection{\ref{sec:cond_TNTI}}, we will compare and contrast the conditional statistics in the outer region, obtained based on detection of the TNTI, between the ZPG and APG TBL cases.

\begin{figure}[t!]
\centering
\includegraphics[width=0.47\linewidth]{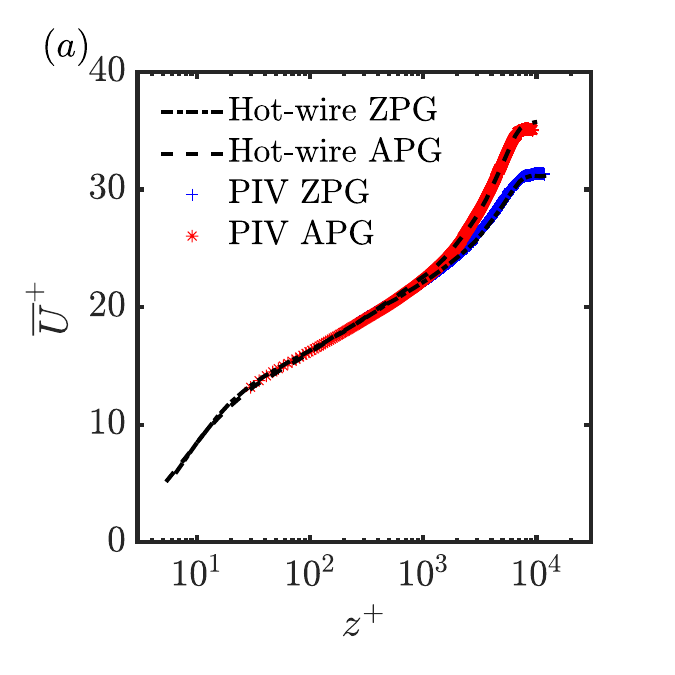}
\includegraphics[width=0.47\linewidth]{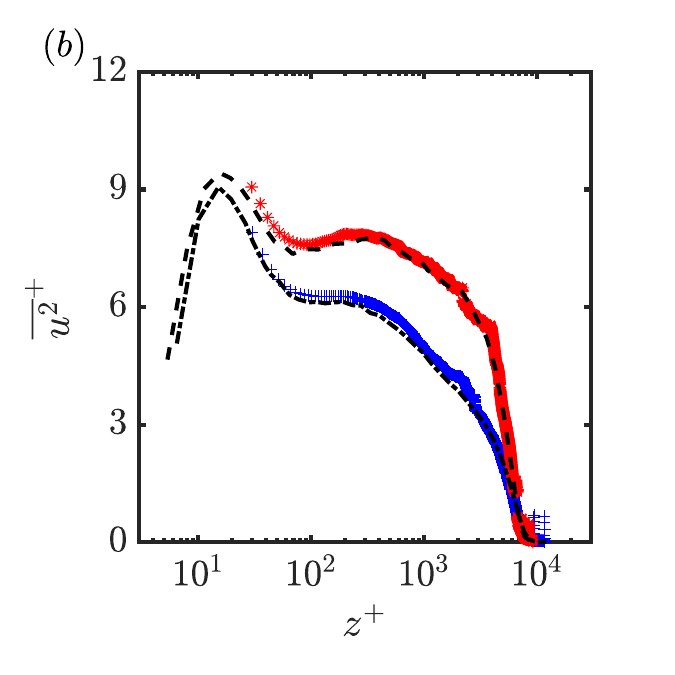}
\caption{Mean viscous scaled  (\textit{a}) streamwise velocity $U^+$ and (\textit{b}) streamwise velocity fluctuations $\overline{u^2}^+$ for the ZPG and APG cases. The dash-dotted and dashed lines  lines in (\textit{a}) and (\textit{b}) are data for ZPG and APG cases obtained from the hot-wire performed in Deshpande\textit{ et al}.\cite{deshpande2023reynolds} at comparable Reynolds number and pressure gradient of the current PIV measurement.}
\label{fig:vel_prof}
\end{figure}

\subsection{Measurement validation} \label{sec:vel_profiles}
Viscous-scaled mean streamwise velocity profiles ${\overline{U}^+ = \overline{U}/U_\tau}$ for the ZPG and APG experiments are plotted in figure~\ref{fig:vel_prof}(\textit{a}). 
Here, the over-line denotes averaging in time (for profiles from hotwire data) and ensemble averaging (for PIV). 
In our current narrow streamwise FOV PIV measurement, we observed negligible development of the TBL along the streamwise ($x$) direction, owing to which there were negligible changes in the turbulent statistics. 
Hence, all the PIV profiles plotted henceforth have been obtained based on averaging across the $x-$direction as well as across the number of images. 
Profiles of the hot-wire measurements by Deshpande\textit{ et al}.\cite{deshpande2023reynolds}, for ZPG and APG under comparable conditions to the current PIV data, are included in figure~\ref{fig:vel_prof}(\textit{a}) for comparison purposes. It is noteworthy that the friction velocity $U_\tau$ for each case is determined from the OFI measurement conducted at the same $x-$location as the PIV measurement, which has been discussed previously in \textsection{\ref{ofi}}.

Figure~\ref{fig:vel_prof}(\textit{a}) illustrates that both the ZPG and APG experiment profiles exhibit good agreement with the hot-wire results. The first reliable data point of the current PIV measurement in the vicinity of the wall is at ${z^+ \approx 30}$ (${z\approx 1100}$~\textmu m for the ZPG and 1400~\textmu m for the APG cases), below which the reflection from the wall hinders accurate estimation. 
Additionally, figure~\ref{fig:vel_prof}(\textit{a}) demonstrates that $\overline{U}^+$ for both ZPG and APG overlap reasonably well up to ${z^+ \lesssim 750}$, beyond which the $\overline{U}^+$ of the APG becomes greater than that of the ZPG in the wake region \cite{perry1995wall,harun2013pressure,knopp2021experimental,deshpande2023reynolds}.

The non-dimensionalized mean streamwise velocity fluctuations ${\overline{u^2}^+ = \overline{u^2}/U_\tau^2}$ for both the ZPG and APG, along with the hot-wire measurements, are shown in figure~\ref{fig:vel_prof}(\textit{b}). This plot demonstrates favorable agreement between the current PIV data and the hot-wire results. As anticipated based on the literature \cite{perry1995wall,harun2013pressure,deshpande2023reynolds}, the $\overline{u^2}^+$ profile of the APG case exceeds that of the ZPG in the outer region of the TBL, despite the nominally similar $Re_{\tau}$ for the two cases.
This is physical and is attributed to the energization of the inertial scales/eddies on imposition of the APG, and is not an artefact of $U_{\tau}$-scaling \cite{harun2013pressure}.
This comparison between PIV and hotwire validates the former and paves the way for deeper analysis into the TNTI flow physics in the next subsections.


\subsection{TNTI detection}\label{sec:TNTI_detect}
As discussed in the introduction, various methods have been proposed in the literature for detecting the TNTI. Given the high resolution of our current PIV data, we opted to employ the local kinetic energy (LKE) method, as previously demonstrated by Chauhan et al.~\cite{chauhan_entrainment_2014} and discussed in the context of Eq.~\eqref{LKE}. This method relies on a threshold parameter $k_{th}$, where turbulent regions are identified when ${\tilde{k} \gtrsim k_{th}}$ and non-turbulent regions are labelled when ${\tilde{k} < k_{th}}$. 
To choose the threshold value for the present investigation, we followed the same procedure as done by Chauhan et al. \cite{chauhan_entrainment_2014}, of selecting $k_{th}$ based on comparison with the intermittency profiles obtained from reference hotwire measurement. 
We found $k_{th} = 0.2$ to yield satisfactory agreement between the intermittency profiles derived from the PIV data and those obtained from independent hot-wire data, although the comparison is not presented here.

\begin{figure}[t!]
\centering
\includegraphics[width=0.75\linewidth]{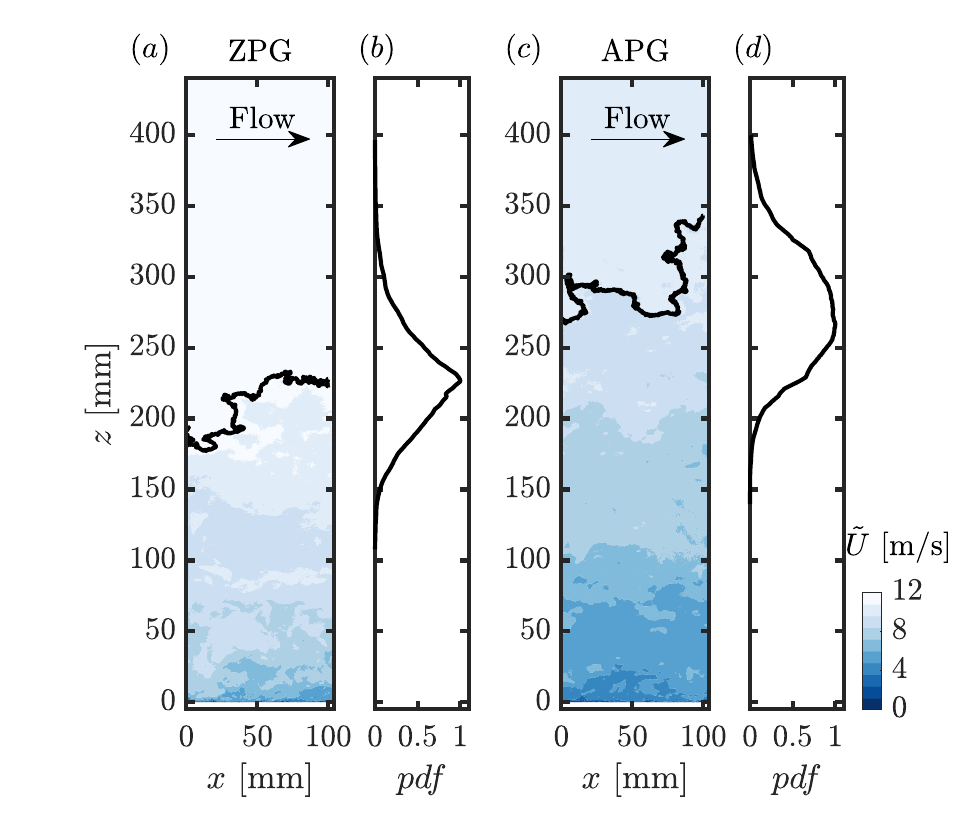}
\caption{Instantaneous streamwise velocity $\tilde{U}$ in an $xz-$plane for (\textit{a}) ZPG and (\textit{c}) APG cases. (\textit{b}) and (\textit{d}) display the probability density function ($pdf$) of the detected TNTI for ZPG and APG cases, respectively. The black lines in (\textit{a}) and (\textit{c}) represent the instantaneous TNTI detected for each corresponding case. In these plots, the flow direction is from left to right, as indicated by the arrow in (\textit{a}) and (\textit{c}).}
\label{fig3}
\end{figure}

Next, we use this criterion to determine the local interface position for both the ZPG and APG cases. 
Samples of the interface are illustrated in figure~\ref{fig3}(\textit{a}) and (\textit{c}) for the ZPG and APG experiments, respectively. 
The instantaneous streamwise velocity ($\tilde{U}$) associated with the detected interface are also provided in these figures. 
For the ZPG case in figure~\ref{fig3}(\textit{a}), the interface appears to vary within the range of ${180~\mathrm{mm} \lesssim z \lesssim 230~\mathrm{mm}}$ for this particular flow realization. 
On the other hand, the interface appears to vary within ${270~\mathrm{mm} \lesssim z \lesssim 350~\mathrm{mm}}$ in case of the APG, clearly indicating greater thickness of the TBL in case of the APG (which is also expected).

The above discussion (and figure \ref{fig3}) not only suggest a greater TBL thickness for APG TBLs but also a greater wall-normal extent of the TNTI region, as compared to the ZPG TBL case. 
To investigate this further, we analyze the probability density function ($pdf$) of the TNTI across the 2000 PIV fields acquired for both TBL cases. 
Figures~\ref{fig3}(\textit{b,d}) depict this TNTI for the ZPG and APG TBL, respectively.
In case of the ZPG TBL, the $pdf$ peaks at approximately ${z \approx 227}$~mm (${z^+ \approx 6300}$) and exhibits a relatively narrow distribution. 
In contrast, the $pdf$ for the APG case peaks at $z$ $\sim$ 270\:mm and clearly exhibits a broader variation compared to the ZPG TBL. 
This analysis makes it imperative to investigate whether the conditional statistics based on the TNTI detection affect the APG TBL statistics more significantly than that in case of ZPG TBLs.
We use this as a motivation to investigate the conditionally averaged turbulence statistics in the next subsection.

\subsection{Conditional statistics based on detection of TNTI}\label{sec:cond_TNTI}
\begin{figure}[t!]
\centering
\includegraphics[width=0.8\linewidth]{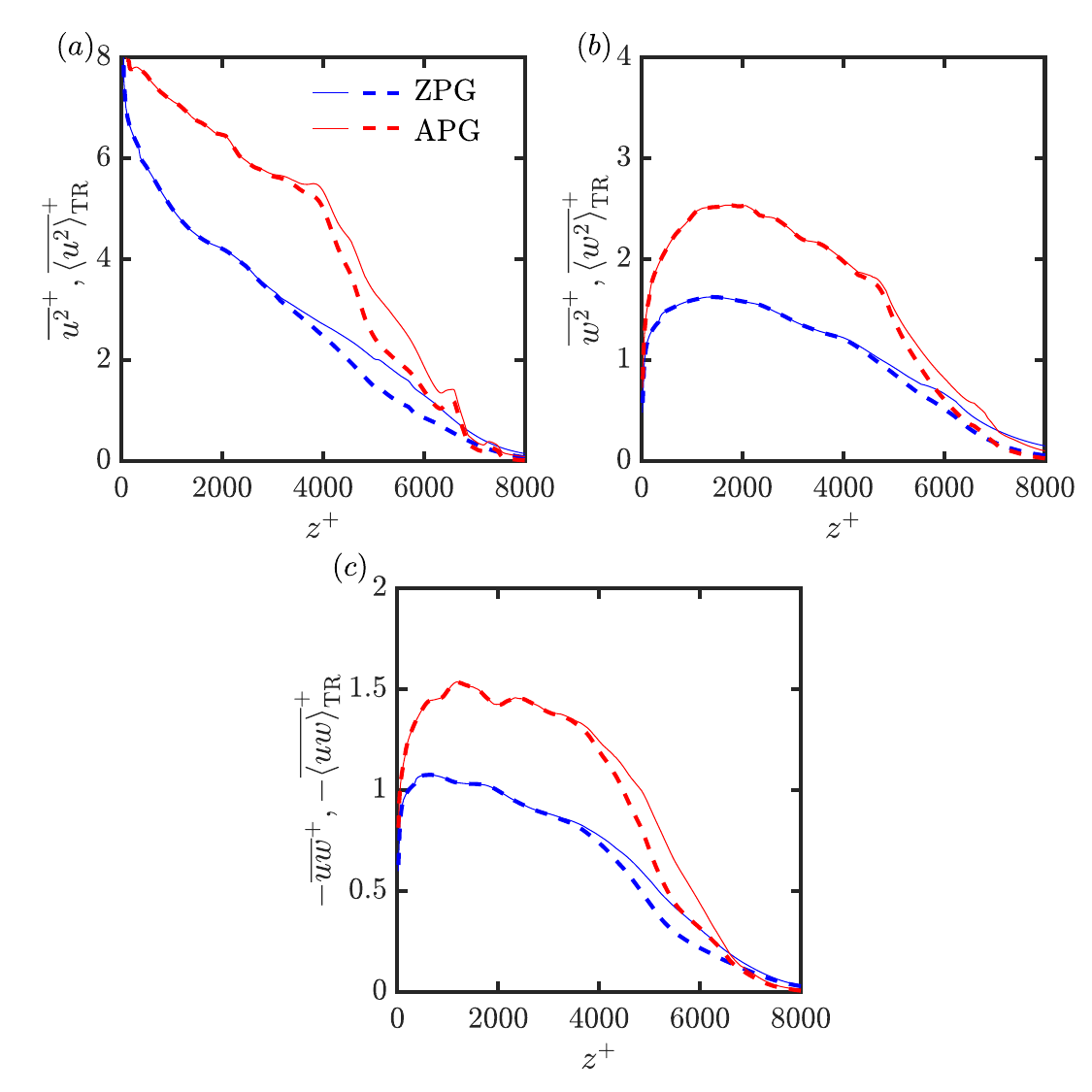}
\caption{Viscous-scaled (\textit{a}) streamwise ($\overline{u^2}^+$), and (\textit{b}) wall-normal variances ($\overline{w^2}^+$), and (\textit{c}) Reynolds shear stresses ($\overline{uw}^+$) for the ZPG (blue lines) and APG cases (red lines). 
The solid lines represent ensemble-averaged data in space and across samples, with fluctuations estimated based on the conventional Reynolds decomposition (\emph{i.e.}, accounting for both the turbulent region and the external freestream flow).
While the dashed lines depict conditionally averaged data after identification of the turbulent region. 
The latter is denoted as, for example, $\overline{\langle {u^2}\rangle}_{\mathrm{TR}}^+$.}
\label{fig5}
\end{figure}

To illustrate the bias in mean turbulence quantities obtained based on the conventional Reynolds decomposition (\emph{i.e.}, when both turbulent and non-turbulent regions are included), we estimate turbulence statistics obtained through conditional averaging below the detected TNTI interface, \emph{i.e.} solely based on the turbulent region. 
The comparison between these two statistics is shown in figure~\ref{fig5}, where conditional averaging within the turbulent region is denoted by angle brackets as ${\langle . \rangle_{\mathrm{TR}}}$, with the subscript `TR' indicating turbulent region.

The mean viscous-scaled streamwise $\overline{u^2}^+$, wall-normal $\overline{w^2}^+$ velocity fluctuations, and Reynolds shear stresses $\overline{uw}^+$ for both the ZPG and APG cases are presented in figure~\ref{fig5}. 
It is evident that the conditionally averaged mean streamwise velocity fluctuations $\overline{\langle {u^2}\rangle}_{\mathrm{TR}}^+$ for the ZPG measurements in figure~\ref{fig5}(\textit{a}) overlap with the mean variance in the region close to the wall (${z^+\lesssim 3000}$), where the interface is not expected (refer $pdf$ plot in figure \ref{fig3}b). 
However, beyond this region (${z^+ > 3000}$), $\overline{\langle {u^2}\rangle}_{\mathrm{TR}}^+$ appears to be reduced compared to the mean statistics, $\overline{u^2}^+$. 
This observation aligns with previous findings by Reuther \& K{\"a}hler~\cite{reuther2020effect}, who attributed the higher magnitude of $\overline{u^2}^+$ away from the wall, to the bias introduced by including non-turbulent regions in the averaging process. 
A similar trend can also be observed for $\overline{w^2}^+$ and $\overline{uw}^+$ in figures~\ref{fig5}(\textit{b}) and \ref{fig5}(\textit{c}) for the ZPG case, where the conditionally averaged statistics are smaller than those of the mean statistics away from the wall.

Figure~\ref{fig5} also illustrates that the same observation holds true for the APG case as for the ZPG case, with the conditionally averaged statistics appearing unaffected close to the wall and reduced compared to the mean statistics away from the wall. 
However, owing to the larger magnitude of turbulence statistics in the APG case, the reduction in magnitude of the conditional statistics is much more pronounced compared to the ZPG case. 
Interestingly, the wall-normal height at which the interface begins to influence the conditionally averaged statistics (in case of APG TBL) is comparable to that observed for the ZPG case, when compared in viscous units (at ${z^+\gtrsim 3000}$).
This is despite the broader variations of the interface seen in the APG case, as shown previously in figure~\ref{fig3}(\textit{d}). 
Further investigations are needed to comprehensively quantify the influence of the interface on higher-order turbulence statistics in an APG TBL, and compare them with ZPG TBL.

\section{Concluding remarks and future work}

The present study reports well-resolved 2-D PIV experiments in high Reynolds number ($Re_{\tau}$ $\gtrsim$ 7000) turbulent boundary layers.
High resolution is obtained in the wall-normal direction by vertically stacking PIV cameras in a `tower' configuration, to resolve the wall-normal variation of the turbulence statistics for comparison between a ZPG and APG TBL scenario.
The turbulence statistics from PIV are validated against previously published hot-wire measurements, which were acquired at similar flow conditions to those considered for the PIV.

The high wall-normal resolution of the PIV measurements makes them ideally suited for analysis of the turbulent/non-turbulent interface (TNTI) in a high $Re_{\tau}$ TBL. 
To this end, the previously established methodology of Chauhan et al. \cite{chauhan_entrainment_2014} is deployed to compare the TNTI for a high $Re_{\tau}$ ZPG and APG TBL.
On comparing ZPG and APG TBLs at nominally same $Re_{\tau}$, it is found that the TNTI interface exists across a greater physical wall-normal distance in case of APG TBL as compared to ZPG TBL.
Finally, the detection of the TNTI interface is used to estimate and compare conditional statistics for both TBL cases.
In general, the conditionally-averaged turbulence statistics are found to be lower than the conventionally obtained mean statistics, based on the Reynolds decomposition, for both the TBL cases.
However, owing to the significantly energetic outer region in an APG TBL, this drop in magnitude of the conditional statistics is much more significant than that noted for ZPG TBL.

Future work will focus on testing various TNTI detection methods and detailed comparisons between the TNTI of a ZPG and APG TBL. 
Further investigations of the conditionally-averaged statistics for higher order moments can also provide further insight into the TNTI flow physics.

\section{Acknowledgements}

I.M. and R.D. gratefully acknowledge funding from the US Office of Naval Research (ONR) and ONR Global program\# N62909-23-1-2068. I.M. also acknowledges funding from the Australian Research Council.
R. D. acknowledges support from University of Melbourne's Postdoctoral Fellowship.

%
%


\begin{thebibliography}{6}
%


\bibitem{daSilva2014}
da Silva, C.B., Hunt, J.C.R., Eames, I., Westerweel, J.: Interfacial layers between regions of different turbulence intensity. \textit{Annual Review of Fluid Mechanics}, 46, 567--590 (2014).

\bibitem{clauser1954turbulent}
Clauser, F. H.: Turbulent boundary layers in adverse pressure gradients. \textit{Journal of the Aeronautical Sciences}, 21(2), 91--108 (1954).

\bibitem{harun2013pressure}
Harun, Z., Monty, J. P., Mathis, R., Marusic, I.: Pressure gradient effects on the large-scale structure of turbulent boundary layers. \textit{Journal of Fluid Mechanics}, 715, 477--498 (2013). \url{https://doi.org/10.1017/jfm.2012.579}

\bibitem{knopp2021experimental}
Knopp, T., Reuther, N., Novara, M., Schanz, D., Sch{\"u}lein, E., Schr{\"o}der, A., K{\"a}hler, C. J.: Experimental analysis of the log law at adverse pressure gradient. \textit{Journal of Fluid Mechanics}, 918 (2021). \url{https://doi.org/10.1017/jfm.2021.296}


\bibitem{marusic2015evolution}
Marusic, I., Chauhan, K. A., Kulandaivelu, V., Hutchins, N.: Evolution of zero-pressure-gradient boundary layers from different tripping conditions. \textit{Journal of Fluid Mechanics}, 783, 379--411 (2015). \url{https://doi.org/10.1017/jfm.2015.501}

\bibitem{bross2019interaction}
Bross, M., Fuchs, T., K{\"a}hler, C. J.: Interaction of coherent flow structures in adverse pressure gradient turbulent boundary layers. \textit{Journal of Fluid Mechanics}, 873, 287--321 (2019). \url{https://doi.org/10.1017/jfm.2019.130}

\bibitem{song2002effects}
Song, S., Eaton, J.: The effects of wall roughness on the separated flow over a smoothly contoured ramp. \textit{Experiments in Fluids}, 33(1), 38--46 (2002). \url{https://doi.org/10.1007/s003480100307}

\bibitem{perry1995wall}
Perry, A. E., Maru{\v{s}}i{\'c}, I.: A wall-wake model for the turbulence structure of boundary layers. Part 1. Extension of the attached eddy hypothesis. \textit{Journal of Fluid Mechanics}, 298, 361--388 (1995). \url{https://doi.org/10.1017/S002211209500128X}

\bibitem{monty2011parametric}
Monty, J. P., Harun, Z., Marusic, I.: A parametric study of adverse pressure gradient turbulent boundary layers. \textit{International Journal of Heat and Fluid Flow}, 32(3), 575--585 (2011). \url{https://doi.org/10.1016/j.ijheatfluidflow.2011.02.006}

\bibitem{henry2007}
Ng, H. C., Marusic, I., Monty, J. P., Hutchins, N., Chong, M. S.: Oil film interferometry in high Reynolds number turbulent boundary layers. \textit{Proceedings of the $16^{th}$ Australasian Fluid Mechanics Conference} (2007).

\bibitem{marusic2010}
Marusic I., McKeon B. J., Monkewitz P. A., Nagib H. M., Smits A. J., Sreenivasan K. R.: Wall-bounded turbulent flows at high Reynolds numbers: Recent advances and key issues. \textit{Physics of Fluids}, 22(6), 1 (2010).

\bibitem{fernholz1996}
Fernholz H. H., Janke G., Schober M., Wagner P. M., Warnack D.: New developments and applications of skin-friction measuring techniques. \textit{Measurement Science and Technology}.1, 7(10):1396, (1996).





\bibitem{bobke2017history}
Bobke, A., Vinuesa, R., {\"O}rl{\"u}, R., Schlatter, P.: History effects and near equilibrium in adverse-pressure-gradient turbulent boundary layers. \textit{Journal of Fluid Mechanics}, 820, 667--692 (2017). \url{https://doi.org/10.1017/jfm.2017.249}






\bibitem{squire_morrill-winter_hutchins_schultz_klewicki_marusic_2016}
Squire, D. T., Morrill-Winter, C., Hutchins, N., Schultz, M. P., Klewicki, J. C., Marusic, I.: Comparison of turbulent boundary layers over smooth and rough surfaces up to high Reynolds numbers. \textit{Journal of Fluid Mechanics}, 795 (2016). \url{https://doi.org/10.1017/jfm.2016.196}



\bibitem{deshpande2023reynolds}
Deshpande, R., van den Bogaard, A., Vinuesa, R., Lindić, L., Marusic, I.: Reynolds-number effects on the outer region of adverse-pressure-gradient turbulent boundary layers. (2023). 
\textit{Physical Review Fluids},{8}(12),124604.
\url{https://doi.org/10.1063/5.0010167}

\bibitem{abu2020effect}
Rowin W.A. and Ghaemi S.: Effect of Reynolds number on turbulent channel flow over a superhydrophobic surface.
\textit{Physics of Fluids},{32}(7),(2020).
\url{https://doi.org/10.1063/5.0010167}

\bibitem{reuther2020effect}
Reuther, N. and K{\"a}hler, C.J., 2020. Effect of the intermittency dynamics on single and multipoint statistics of turbulent boundary layers. Journal of Fluid Mechanics, 897, p.A11.



  
\bibitem{corrsin1955free}
  Corrsin, S. and Kistler, A.L.:Free-stream boundaries of turbulent flows,
  \textit{NASA},
  {1955}
\bibitem{yang2020turbulent}
  Yang, J. and Yoon, M. and Sung, H.J.: The turbulent/non-turbulent interface in an adverse pressure gradient turbulent boundary layer,
\textit{International Journal of Heat and Fluid Flow},
  {86},
  {108704},
  {2020},
  {Elsevier}

\bibitem{chauhan2014scaling}
  Chauhan, K. and Philip, J. and Marusic, I.: Scaling of the turbulent/non-turbulent interface in boundary layers,  
  \textit{Journal of Fluid Mechanics},
  {751},
  {298--328},
  {2014},
  {Cambridge University Press}

\bibitem{chauhan_entrainment_2014}
  Chauhan, K., Philip, J., de Silva C. M., Hutchins, N., Marusic, I., The turbulent/non-turbulent interface and entrainment in a boundary layer,  
  \textit{Journal of Fluid Mechanics},
  {vol. 742},
  {pp. 119-151},
  {2014}
  
\bibitem{de_wavelet_2023}
  De, S., Anand, A., Diwan, S. S., A wavelet-based detector function for characterizing intermittent velocity signals,  
  \textit{Experiments in Fluids},
  {64:180},
  {2023}
  
\bibitem{chen_similarity_2024}
  Chen, L., Tang, Z., Fan, Z., Jiang, N., Outer-layer self-similarity of the turbulent boundary layer based on the turbulent/non-turbulent interface,  
  \textit{Physical Review Fluids 9},
  {034607},
  {2024}
  
\bibitem{semin_superlayer_2011}
  Semin, N. V., Golub, V. V., Elsinga, G. E., Westerweel, J., Laminar superlayer in a turbulent boundary layer,  
  \textit{Technical Physical Letters},
  {vol. 37},
  {no. 12},
  {pp. 1154-1157},
  {2011}
  
\bibitem{fransson_transition_2005}
  Fransson, J. H. M., Matsubara M., Alfredsson, P. H., Transition induced by free-stream turbulence,  
  \textit{Journal of Fluid Mechanics},
  {vol. 527},
  {pp. 1-25},
  {2005}
  
\bibitem{hedley_properties_1974}
  Hedley, T. B. \& Keffer, J. F., Some turbulent/non-turbulent properties of the outer intermittent region of a boundary layer,  
  \textit{Journal of Fluid Mechanics},
  {vol. 64},
  {pp. 645-678},
  {1974}
  
\bibitem{hedley_decisions_1974}
  Hedley, T. B. \& Keffer, J. F., Turbulent/non-turbulent decisions in an intermittent flow,  
  \textit{Journal of Fluid Mechanics},
  {vol. 64},
  {pp. 625-644},
  {1974}

\bibitem{buxton_entrainment_2023}
  Buxton, O. R. H. \& Chen, J., The relative efficiencies of the entrainment of mass, momentum, and kinetic energy from a turbulent background,  
  \textit{Journal of Fluid Mechanics},
  {vol. 977},
  {R2},
  {2023}

\bibitem{krug_2017}
Krug, D., Philip J., Marusic I., Revisiting the law of the wake in wall turbulence. \textit{Journal of Fluid Mechanics},
  {vol. 811},
  {421-35},
  {2017}.

\end{thebibliography}
\end{document}